\documentclass[aps,prb,a4paper,amsmath,showpacs,floatfix,twocolumn]{revtex4}
\usepackage{graphicx}
\def\lsim{\mathrel {\vcenter {\baselineskip 0pt \kern 0pt
    \hbox{$<$} \kern 0pt \hbox{$\sim$} }}}

\begin{document}

\title{Stability of smectic phases in hard-rod mixtures}

\author{Yuri Mart\'{\i}nez-Rat\'on}
\email{yuri@math.uc3m.es}

\affiliation{
Grupo Interdisciplinar de Sistemas Complejos (GISC), Departamento de Matem\'aticas, Escuela Polit\'ecnica Superior, Universidad Carlos III de Madrid, Avenida de la Universidad 30, E-28911 Legan\'es, Madrid, Spain} 

\author{Enrique Velasco}
\email{enrique.velasco@uam.es}
\affiliation{Departamento de F\'{\i}sica Te\'orica de la Materia Condensada
and Instituto de Ciencia de Materiales Nicol\'as Cabrera,
Universidad Aut\'onoma de Madrid, E-28049 Madrid, Spain.}

\author{Luis Mederos}
\email{lmederos@icmm.csic.es}
\affiliation{
Instituto de Ciencia de Materiales, Consejo Superior de Investigaciones Cient\'{\i}ficas, E-28049 Cantoblanco, Madrid, Spain}

\date{\today}

\begin{abstract}
Using density-functional theory, we have analyzed the phase behavior of binary
mixtures of hard rods of different lengths and diameters. Previous
studies have shown a strong tendency of smectic phases of these mixtures 
to segregate and, in some circumstances, to form microsegregated phases. 
Our focus in the present work is on the formation of columnar
phases which some studies, under some approximations, have shown to become 
thermodynamically stable prior to crystallization. Specifically we focus
on the relative stability between smectic and columnar phases, a question not 
fully addressed in previous work. Our analysis
is based on two complementary perspectives: on the one hand, an extended 
Onsager theory, which includes the full orientational degrees of freedom but
with spatial and orientational correlations being treated in an 
approximate manner;
on the other hand, we formulate a Zwanzig approximation of fundamental-measure
theory on hard parallelepipeds, whereby orientations
are restricted to be only along three mutually orthogonal axes, but correlations 
are faithfully represented. In the latter case novel, complete phase 
diagrams containing regions of stability of liquid-crystalline phases are calculated. 
Our findings indicate that the restricted-orientation
approximation enhances the stability of columnar phases so as to 
preempt smectic order completely while, in the framework of the
extended Onsager model, with full orientational degrees of freedom
taken into account, columnar phases may preempt a large region of 
smectic stability in some mixtures, but some smectic order still persists.
\end{abstract}
% 64.70.Md  Transitions in liquid crystals
% 64.75.+g  Solubility, segregation, and mixing; phase separation
% 61.20.Gy  Theory and models of liquid structure

\maketitle

\section{Introduction}
Recent work has shown that strong segregation effects arise 
in smectic (S) phases of hard rods of equal diameter but different lengths
when the two components are mixed in varying
proportions. In seminal work Koda and Kimura\cite{Koda} analyzed a 
binary mixture of parallel hard cylinders using Onsager approximation, and
applied a stability analysis. They found two types of S phases: one in which 
layers are identical and contain a mixture of both types of particles, and 
another, microsegregated phase, 
which consists of alternating layers of different compositions. More recent work
has focused on more general mixtures. For example, 
van Roij and Mulder\cite{vanRoij} studied a mixture of parallel cylinders
of different diameters using Onsager theory and a more complete 
bifurcation analysis. They observed that, for
length ratios higher than 1:5, N-N segregation (where N stands for nematic
phase) preempts the N-S and N-C phase transitions (C stands for columnar phase).
%Recent work has shown that interesting microsegregation effects arise
%when hard rods of equal diameter but different length are mixed in varying
%proportions. One particular case of such a mixture, where one of the 
%components consists of hard spheres, was examined by Frenkel and .... 
%In some r\'egime of length-to-breadth ratio for the hard rods, and in the
%limiting case where the latter were all taken to be parallel, the 
%stability of the smectic phase was found to be enhanced with respect 
%to the nematic phase, an effect which can be traced back to a more
%favourable packing when rods are arranged in layers. However, the 
%structure of the smectic phase consists of layers of rods alternating 
%with layers of spheres, in what can be considered as a microsegregated
%phase. Additionally, a macroscopic segregation takes place between
%such a phase and a smectic phase consisting almost entirely of rods.
More recent work by Cinacchi et al.\cite{Cinacchi,Cinacchi1} 
used an extended Onsager theory (EOT) on systems of hard spherocylinders (HSPCs),
but lifting the restriction on orientations, and complete phase diagrams
were obtained containing I (isotropic), N, and S, phases. 
As was expected, a strong tendency toward segregation was observed,
and microsegregated phases did also arise in the model, consisting of
layers made up of the component in larger proportion --the shorter
rods-- alternating with weakly populated and highly interpenetrated
layers of the longer rods. 
%In no case was the symmetric phase
%(layers of long rods alternating with highly interpenetrated
%layers of short rods) observed, nor were phases of alternating layers
%of the two components in proportions close to 50\%. Also, even though
%not excluded in the theoretical model, particles were always seen 
%to lie parallel to the director\cite{...}. 

In the work by Cinacchi et al. the only spatially ordered phase considered
was the S phase. C phases are very common in nature, even in systems composed
of prolate molecules. A simple, one-component model is known not to exhibit
C ordering, which is preempted by the smectic phase. However, the C phase
can be stabilized prior to crystallization in mixtures of HSPC particles with 
orientations restricted to be parallel, as shown by Stroobants et 
al.\cite{Stroobants} using computer 
simulation. Bohle et al.,\cite{Bohle} using a density functional based on the 
weighted-density approximation, considered the effect of polydispersity on the 
stabilization of ordered phases in systems of parallel hard rods. Their 
conclusion is similar to that obtained previously by Bates and Frenkel,
\cite{Bates} {\it viz.},
increasing the degree of polydispersity stabilizes the C phase with
respect to the S phase. An identical behavior was observed in mixtures
with perfectly bimodal length distributions\cite{Bohle} 
(i.e., two-component mixtures).

It is intuitively clear that orientational fluctuations, not taken 
into account in all of these studies, will tend to destroy columnar order, but the 
extent to which this statement is true in mixtures of hard rods is still unknown. 
As a consequence, a general concern is the question of 
whether phases of lower symmetry, C and crystal, may preempt all the structures
observed in preliminary studies; in particular, all previous studies
have shown the C phase to be stable for systems of perfectly parallel rods,
preempting the occurrence of the smectic phase for sufficiently large
length ratios. 
Also, in connection with the work by Cinacchi et al.,\cite{Cinacchi}
a number of questions remain unanswered, particularly in relation with
the theoretical model (EOT), its particular 
implementation, and the approximations used (parametrization of the
distribution functions, etc.). These are the aims of the present paper.
In this paper we have not considered the occurrence
of crystal phases, largely for two reasons: one is that in mixtures
the crystal phase will certainly tend to appear at relatively high packing
fractions, due to packing constraints, which will leave a large region
where other spatially ordered phases may appear. Another reason is
that the EOT is not well suited to study crystallization. Therefore
we will limit ourselves
to studying the instability against general spatial fluctuations of the
N phase.

In the present paper we address the above questions from two 
perspectives. One is the calculation, within EOT,
of the line where the nematic
phase of the mixture gets unstable with respect to columnar-phase
fluctuations. This line sets an approximate upper limit for the region
where smectic phases should be stable. The other is the numerical 
minimization of a fundamental-measure functional as applied to systems 
of hard rods of square 
transversal section [hard parallelepipeds (HPs)] in the Zwanzig approximation,
i.e., considering that the rods can assume only three possible
orientations, namely, along the Cartesian axes $xyz$. In this
way, by using a completely different approach which possesses some advantages
(better treatment of hard-core correlations and exact formulation
of excluded-volume effects) and some disadvantages (restricted
orientations), we may verify whether the predictions of Onsager theory are
artifacts of the theory or of its particular implementation.
We will see that some of the predictions of the two
models agree, which gives us some confidence as to the realistic
nature of the results. On the other hand, the results from the 
fundamental-measure theory 
(FMT) predict
new structures, not observed in the Onsager theory, which poses
interesting new questions and opens up new avenues of research --of
course the extent to which these structures are artifacts of the
Zwanzig approximation or peculiarities of the particle's shape is 
as yet unknown.

In Sec. II we briefly remind the reader the ingredients of the
two theoretical approaches. It is also devoted to present
the methodology used to locate the spinodal line for C ordering 
in the context of the above density-functional theories.
In Sec. \ref{Results} we present a few results pertaining to the
EOT used previously\cite{Cinacchi} to analyze
phase behavior in HSPC mixtures, summarizing relevant features that
appear in the phase diagrams of a few selected mixtures. Complete
phase diagrams, containing I, N, and S phases, are presented. Our purpose
is to compare the corresponding results with those obtained using 
the Zwanzig approximation of FMT as applied to HP particles
and for equivalent mixtures. Density and order-parameter
profiles are shown to better demonstrate the structure of the 
spatially ordered S phases. Also, spinodal lines 
signaling the instability against C-type fluctuations are shown.
The location of this line in connection with the stability of the S phases
is discussed. We end in Sec. \ref{Conclusions}
with a summary of the results and
some final conclusions drawn from the comparison of the results
obtained with the two theoretical approaches.

\section{Theoretical approaches}

In this section we use two versions of the density-functional theory,
a modified Onsager theory and fundamental-measure theory in the
Zwanzig approximation, to obtain phase diagrams of a few selected
mixtures of hard rods. Attention is restricted to spatially disordered
phases [isotropic (I) and nematic (N)] and also to smectic (S) phases.
The more ordered columnar (C) phase will be considered in Sec.
II C.

\subsection{Extended Onsager theory for mixtures of hard spherocylinders}
\label{Onsager}

We analyze the phase behavior of binary mixtures of hard spherocylinders 
(HSPs) of length $L_s$ ($s=1,2$) and the same diameter $\sigma$.
The one-particle
distribution functions of each component will be denoted by 
$\rho_s({\bf r},\hat{\bf\Omega})$. The density functional for the Helmholtz 
free energy ${\cal F}$ is written, as usual, as a sum of ideal 
${\cal F}_{\rm id}$ and excess ${\cal F}_{\rm ex}$ terms:
\begin{equation}
{\cal F}[\{\rho_s\}]={\cal F}_{\rm id}[\{\rho_s\}]+
{\cal F}_{\rm ex}[\{\rho_s\}],
\label{1}
\end{equation}
with the ideal part given by
\begin{equation}
\beta{\cal F}_{\rm id}[\{\rho_s\}]=\sum_{s=1}^2
\int\!\!\int d{\bf r}d\hat{\bf\Omega}\rho_s({\bf r},\hat{\bf\Omega})
\left\{\ln{\rho_s({\bf r},\hat{\bf\Omega})}-1\right\},
\label{2}
\end{equation}
where $\beta=1/kT$. The excess part is given by
\begin{equation}
\beta{\cal F}_{\rm ex}[\{\rho_s\}]=\int_V d{\bf r} 
\Phi_{\rm{ex}}({\bf r};\{\rho_s\}),
\label{3}
\end{equation}
where the excess free-energy density per unit thermal energy, 
$\Phi_{\rm{ex}}({\bf r};\{\rho_s\})$, is written, in the EOT, as
\begin{eqnarray}
\Phi_{\rm{ex}}({\bf r};\{\rho_s\})&=&
\Psi(\eta)\sum_{s=1}^2\sum_{t=1}^2
\int_V d{\bf r}^{\prime}
\int\!\!\!\int d\hat{\bf\Omega}
d\hat{\bf\Omega}^{\prime}\nonumber\\
&\times&\rho_s({\bf r},\hat{\bf\Omega})
\rho_t({\bf r}^{\prime},\hat{\bf\Omega}^{\prime})
f_{st}({\bf r}-{\bf r}^{\prime},\hat{\bf\Omega},\hat{\bf\Omega}^{\prime}).
\label{4}
\end{eqnarray}
$f_{st}$ are overlap functions for the three different interactions
(unity if particles overlap and zero otherwise), and $\Psi(\eta)$ is a
prefactor that depends on the mean packing fraction $\eta$ and that
is chosen to make the theory recover the second virial coefficient 
exactly and approximate the remaining virial coefficients in terms
of the second coefficient, in the manner proposed by Parsons\cite{Parsons}
and Lee.\cite{Lee}
Our implementation of the present theory (see details in 
\cite{Cinacchi,Cinacchi1})
involves first using the decoupling approximation, 
\begin{eqnarray}
\rho_s({\bf r},\hat{\bf\Omega})
\approx\rho_s({\bf r})h_s(\hat{\bf\Omega})
\label{decoupling}
\end{eqnarray}
(i.e.,
particle positions and orientations, characterized, respectively, by the
distribution functions $\rho_s$ and $h_s$, are assumed to be decoupled at the level 
of the one-particle distribution functions), and then parametrizing the 
two distribution functions in terms of suitably normalized 
exponential functions:
\begin{eqnarray}
\rho_s({\bf r},\hat{\bf\Omega})
\approx\rho_s({\bf r})h_s(\hat{\bf\Omega})&=&
x_s\rho\left[\displaystyle{\frac{
e^{\displaystyle\lambda_s\cos{q z}}}{I_0(\lambda_s)}}\right]\nonumber\\
&\times&\left[\frac{\displaystyle e^{\displaystyle\Lambda_sP_2(\cos{\theta})}}
{2\pi\displaystyle\int_{-1}^{1}dx e^{\displaystyle\Lambda_sP_2(x)}}\right],
\label{5}
\end{eqnarray}
where $q=2\pi/d$ is the wave number of the smectic undulation with 
period $d$, $\Lambda_s$ are variational parameters describing the
orientational order ($\Lambda_s=0$ in the
I phase and $\Lambda_s\ne 0$ in the N and S phases), 
$\{\lambda_s\}$ are variational parameters describing the positional
order ($\lambda_s=0$ in the N phase),
$x_s$ the molar fractions of species $s$, $\rho$ the mean density, 
and $I_0(x)$ is the modified Bessel function of order 0. 
Equation (\ref{1}) should be minimized with respect to the 
variational parameters $\{\lambda_s\}$, $\{\Lambda_s\}$, and $d$
to find the equilibrium density profiles and the free energy for
particular values of $\rho$ and $x$. The Gibbs free energy 
per particle $g$ is obtained from 
\begin{eqnarray}
g=\frac{F}{N}+\frac{P}{\rho},\hspace{0.4cm}
P=\left(\rho\frac{\partial}{\partial\rho}-1\right)\frac{F}{V},
\end{eqnarray}
where $P$ is the pressure and $F$ is the equilibrium Helmholtz free energy, 
or directly
from an appropriate derivate of the free-energy density functional 
evaluated at the equilibrium profiles.
Applying a Maxwell double-tangent construction on $g(x)$ for fixed
$P$ allows us to find the coexistence values of the different
phases in the usual way and to obtain the phase diagrams in the
pressure--composition plane.

\subsection{Fundamental-measure theory 
for mixtures of hard parallelepipeds in the Zwanzig approximation}

The FMT used here has been described in detail
in Ref. 10, so that only a brief sketch is provided here
in order to show how it can be numerically implemented to calculate
phase diagrams of mixtures.
We consider a system of HPs, i.e., hard rods of 
rectangular shape. We continue to use the notation introduced in Sec. 
II A for the dimensions of the particles; in this case we 
denote by $L_s$ the length of a parallelepiped of species $s$, with
$\sigma_s$ being the breadth of the particles (assumed to have a square
section).

We have used the fundamental-measure functional for HP
in the approximation that orientations are restricted to lie along
three mutually orthogonal directions (Zwanzig approximation), taken
to be the $xyz$ axes.
Since the unit vector $\hat{\bf\Omega}$ only has three possible
orientations, one can think of the one-particle distribution functions 
$\rho_s({\bf r},\hat{\bf\Omega})$ as being expressed in terms of a set
of Dirac-delta functions along the three axes:
\begin{eqnarray}
\rho_s({\bf r},\hat{\bf\Omega})=
\sum_{\mu}\rho_{s\mu}({\bf r})\delta(\hat{\bf\Omega}-\hat{\bf e}_{\mu}),
\label{densities}
\end{eqnarray}
where $\hat{\bf e}_{\mu}$, $\mu=1,2,3$, are unit vectors along the three 
perpendicular directions $xyz$, respectively. Note that this expression
does not assume a decoupling approximation since it implies a
strong mixing of spatial and orientational degrees of freedom.
The coefficients
of these expansions, $\rho_{s\mu}({\bf r})$, are the local density of 
species $s$ parallel to the $\mu$-axe and represent quantitatively the 
orientational order in the system in the Zwanzig approximation. 
The free-energy functional will be given by Eqs. 
(\ref{1})--(\ref{3}) and (\ref{densities}), where
the excess part of the free-energy density in reduced thermal units, 
obtained in Ref. 10, has 
the form 
\begin{eqnarray}
\Phi_{\rm ex}({\bf r};\{\rho_s\})=
-n_0\ln(1-n_3)+\frac{{\bf n}_1\cdot{\bf n}_2}{
1-n_3}+\frac{n_{2x}n_{2y}n_{2z}}{(1-n_3)^2},\nonumber\\
\label{7}
\end{eqnarray}
with the functions $\{n_{\alpha}\}$ ($\alpha=\{0,1x,1y,1z,2x,2y,2z,3\}$)
being weighted densities obtained as 
\begin{eqnarray}
n_{\alpha}({\bf r})=\sum_{s=1}^2\sum_{\mu=1}^3
\int_V d{\bf r}^{\prime}\rho_{s\mu}({\bf r}^{\prime})\omega_{s\mu}^{(\alpha)}
({\bf r}-{\bf r}^{\prime}),
\label{8}
\end{eqnarray}
where $\omega_{s\mu}^{(\alpha)}$ are characteristic functions whose
spatial integrals give the fundamental measures of the particles
(edge length, surface, and volume).
In this case, instead of the parametrization Eq. (\ref{5}), 
used in the EOT, we use a Fourier expansion to represent smectic order:
\begin{eqnarray}
\rho_{s\mu}({\bf r})=x_s\rho\left[\sum_{k=0}^K\alpha_{s\mu}^{(k)}
\cos{qkz}\right]\gamma_{s\mu},
\label{9}
\end{eqnarray}
(this expression is obviously also valid for the N phase),
where $\{\alpha_{s\mu}^{(k)}\}$ are 
Fourier amplitudes (we impose $\alpha_{s\mu}^{(0)}=1$ and 
uniaxial symmetry, reflected in the condition $\alpha_{sx}^{(k)}=
\alpha_{sy}^{(k)}$). The coefficients $\{\gamma_{s\mu}\}$ (which
play the role of $f_s(\hat{\bf\Omega})$ in the Onsager theory) are 
the spatially average probabilities to find 
a particle of species $s$ parallel to the $\mu$ axis (again 
uniaxial symmetry implies $\gamma_{sx}=\gamma_{sy}$).
The number of terms in the expansion, $K+1$, is chosen so that
$\alpha_{s\mu}^{(K)}<10^{-7}$. 

In the nematic limit $\rho_s({\bf r},\hat{\bf\Omega})=
\rho_sh_s(\hat{\bf\Omega})$, and the coefficients $\gamma_{s\mu}$ are easily
obtained in terms of the nematic order parameters $\{Q_s\}$ 
(with $-1/2\le Q_s\le 1$) by imposing the conditions
\begin{eqnarray}
1=\int d\hat{\bf\Omega} h_s(\hat{\bf\Omega}),\hspace{0.4cm}
Q_s=\int d\hat{\bf\Omega} h_s(\hat{\bf\Omega})P_2(\hat{\bf\Omega}
\cdot{\bf e}_z),
\end{eqnarray}
[$P_2(x)$ is the second-order Legendre polynomial] which lead to
\begin{eqnarray}
\gamma_{s\mu}=\frac{1}{3}\left[1+\left(3\delta_{\mu z}-1\right)Q_s\right],
\end{eqnarray}
with $\delta_{\mu\nu}$ the Kronecker delta. We have selected the 
nematic director to be parallel to the $z$ axis. 

With these definitions the weighted densities can be easily calculated 
from Eqs. (\ref{8}) and (\ref{9}) as 
\begin{eqnarray}
n_{\alpha}(z)=\rho\sum_{s\mu}x_s\gamma_{s\mu}\sum_{k=0}^K
\alpha_{s\mu}^{(k)}\hat{\omega}^{(\alpha)}_{s\mu}(qk)\cos(qkz),
\end{eqnarray}
where the functions $\hat{\omega}^{(\alpha)}_{s\mu}$ are the
Fourier transforms of effective one-dimensional weights
resulting from partial integration over $x$ and $y$:
\begin{eqnarray}
&&\hat{\omega}^{(0)}_{s\mu}(qk)=j_0\left(\frac{q}{2}k\sigma_{\mu z}^s
\right),\nonumber\\
&&\hat{\omega}^{(1\nu)}_{s\mu}(qk)=\sigma_{\mu\nu}^s j_{\delta_{\nu z}}\left(
\frac{q}{2}k\sigma_{\mu z}^s\right),\nonumber\\
&&\hat{\omega}^{(2\nu)}_{s\mu}(qk)=\frac{v_s}{\sigma_{\mu\nu}^s}
j_{1-\delta_{\nu z}}\left(\frac{q}{2}k\sigma_{\mu z}^s\right),\nonumber\\
&&\hat{\omega}^{(3)}_{s\mu}(qk)=v_sj_1\left(\frac{q}{2}k\sigma_{\mu z}^s
\right),
\end{eqnarray}
with $v_s=L_s\sigma_s^2$ the particle volumes, 
$\sigma_{\mu\nu}^s=\sigma_s+\left(L_s-\sigma_s\right)\delta_{\mu\nu}$, 
$j_0(x)=\cos x$, and $j_1(x)=\sin x/x$.

The ideal part of the free energy density in reduced thermal units 
for this model is 
\begin{eqnarray}
\Phi_{\rm{id}}(z)=\sum_{s\mu}\rho_{s\mu}(z)
\left[\ln\rho_{s\mu}(z)-1\right].
\end{eqnarray}
Thus the total free energy per unit volume can be calculated as 
\begin{eqnarray}
\frac{\beta{\cal F}}{V}
=\frac{1}{d}\int_0^d dz\left[\Phi_{\rm{id}}(z)+\Phi_{\rm ex}(z)
\right].
\label{phi}
\end{eqnarray}
Now Eq. (\ref{phi}) is minimized with respect to the
parameters $\gamma_{s\mu}$, $\alpha_{s\mu}^{(k)}$, and $d$ to find the 
equilibrium density profiles for fixed $\rho$ and $x$. The pressure can
be evaluated from
\begin{eqnarray}
\beta P&=&\frac{1}{d}\int_0^d dz\Bigg\{
\frac{n_0(z)}{1-n_3(z)}+\frac{{\bf n}_1(z)\cdot {\bf n}_2(z)}
{\left[1-n_3(z)\right]^2}\nonumber\\
&+&2\frac{n_{2x}(z)n_{2y}(z)n_{2z}(z)}{\left[1-n_3(z)\right]^3}\Bigg\},
\label{presion}
\end{eqnarray}
and the phase diagram is obtained as indicated in Sec. \ref{Onsager}.
The results for phase diagrams of different mixtures (including the
one-component limit) pertaining to this
theory, and also to the EOT approach outlined in Sec. II B,
will be presented in Sec. \ref{Results}. Also, density and order-parameter
profiles for the smectic phases will be shown in order to better grasp the
structure of these phases. These profiles are defined as follows. The total
density profile is
\begin{eqnarray}
\rho_s(z)=\sum_{\mu}\rho_{s\mu}(z),
\end{eqnarray}
while the nematic order-parameter profile is defined by
\begin{eqnarray}
Q_s(z)=\frac{1}{2}\left[\frac{3\rho_{sz}(z)}{\rho_s(z)}-1\right].
\end{eqnarray}

\subsection{Instability against spatial fluctuations}
\label{columnar}

In this section we present the formalism used to analyze the instabilities 
against spatial fluctuations.
We first describe the bifurcation analysis 
that we have used to locate the instability points of the N phase against 
spatial ordering. Our aim is to see whether the phase behavior of the 
above HSPC mixtures is strongly modified by the presence of spatially 
ordered phases with a symmetry lower than that of the S phase. In the present 
paper we are interested in locating the spinodal line corresponding to C
ordering, since this is the phase that, following previous work on
parallel hard rods, might become stable prior to crystallization. 

Instabilities against both S and C fluctuations can be explored at the
same time by examining the response function of the N phase
against general spatial fluctuations of a wave vector ${\bf q}$.
The free-energy change $\delta F$ associated with fluctuations $\delta
\rho_s$ in the one-particle distribution functions of the two components
of the mixture can be expressed, up to second order in the fluctuations, 
in terms of the second functional derivative 
of the free-energy functional evaluated at the nematic phase:
\begin{eqnarray}
\delta F&=&\frac{1}{2}\sum_{s=1}^2\sum_{t=1}^2
\int_V d{\bf r}\int d\hat{\bf\Omega}
\int_V d{\bf r}^{\prime}\int d\hat{\bf\Omega}^{\prime}\nonumber\\
&\times&\left.\frac{\delta^2{\cal F}}{\delta\rho_s({\bf r},\hat{\bf\Omega})
\delta\rho_t({\bf r}^{\prime},\hat{\bf\Omega}^{\prime})}\right|_{\rm N}
\delta\rho_s({\bf r},\hat{\bf\Omega})
\delta\rho_t({\bf r}^{\prime},\hat{\bf\Omega}^{\prime}),\nonumber\\
\end{eqnarray}
The instability is signaled by the equation $\delta F=0$.
Two different approaches can be followed here. In one, we assume that
fluctuations are realized at constant mean density $\rho$, so that the
minimization of ${\cal F}$ is performed via a Lagrange multiplier or,
equivalently, by choosing fluctuations $\delta\rho_s$ that conserve
the mean density. This is the approach to be followed in the canonical
ensemble where the composition $x$ and the mean density $\rho$ are 
fixed. In the other approach, general fluctuations, not 
necessarily mean density-conserving, are considered, and the second functional
derivative is most conveniently expressed in terms of the
direct correlation function $C^{}_{st}$ evaluated at the nematic phase,
\begin{eqnarray}
\left.\frac{\delta^2[\beta{\cal F}]}{\delta\rho_s({\bf r},\hat{\bf\Omega})
\delta\rho_t({\bf r}^{\prime},\hat{\bf\Omega}^{\prime})}\right|_{\rm N}
&=&\frac{\delta({\bf r}-
{\bf r}^{\prime})\delta(\hat{\bf\Omega}-\hat{\bf\Omega}^{\prime})\delta_{st}}{\rho_s^{(N)}(\hat{\bf\Omega})}\nonumber \\&-&
C^{}_{st}
({\bf r}-{\bf r}^{\prime},\hat{\bf\Omega},
\hat{\bf\Omega}^{\prime}),
\end{eqnarray}
where $C^{}_{st}$ is the direct correlation function 
evaluated at the nematic phase,
\begin{eqnarray}
C_{st}({\bf r}-{\bf r}^{\prime},\hat{\bf\Omega},\hat{\bf\Omega}^{\prime})=
-\frac{\delta^2 \left[\beta {\cal F}_{\rm ex}\right]}{\delta \rho_{s}({\bf r},\hat{\bf\Omega})
\delta\rho_{t}({\bf r}^{\prime},\hat{\bf\Omega}^{\prime})}\Bigg|_{\rho_{s}({\bf r},\hat{\bf\Omega})=
\rho^{(N)}_{s}(\hat{\bf\Omega})}.\nonumber\\
\hspace*{5.cm} (23) \nonumber
\end{eqnarray}
This route is followed in the calculations using the isobaric ensemble
where the quantities to be fixed are the composition $x$ and the pressure $P$.
Depending on the ensemble used the fluctuation has to be chosen
accordingly. For the Onsager model we have chosen to use the canonical
ensemble; in the spirit of the decoupling approximation, we take
\begin{eqnarray}
\delta\rho_s({\bf r},\hat{\bf\Omega})=\epsilon_se^{i{\bf q}\cdot{\bf r}}
\rho_s^{(N)}(\hat{\bf\Omega})=\epsilon_s\rho_se^{i{\bf q}\cdot{\bf r}}h_s^{(N)}(\hat{\bf\Omega}),
\setcounter{equation}{24} 
\end{eqnarray}
where $h_s^{(N)}$ is the orientational 
distribution function of the nematic phase, ${\bf q}$ an
arbitrary wave vector, and $\epsilon_s$ an amplitude giving the strength of the
fluctuation. The above fluctuation conserves 
the value of the order parameters $\{Q_s\}$ as the mean density is fixed.
The increase in free energy can be written in matrix form as
\begin{eqnarray}
\frac{\delta[\beta{\cal F}]}{V}=
\frac{1}{2}\sum_{s=1}^2\sum_{t=1}^2\epsilon_s\rho_s
{\cal T}_{st}\epsilon_t\rho_t=\frac{1}{2}
{\bf a}^{\rm T}\cdot{\cal T}\cdot{\bf a}
\end{eqnarray}
with ${\bf a}^T=(a_1,a_2)$, and $a_s=\epsilon_s\rho_s$ 
the components of a two-dimensional perturbation vector.
The stability of the nematic phase against spatial fluctuations can be assessed by
examining the signs of the eigenvectors of the $2\times 2-{\cal T}$ matrix, with
\begin{eqnarray}
{\cal T}_{st}=
{\cal T}_{st}(x;{\bf q},\rho)\equiv
\frac{\delta_{st}}{\rho_s}-
\left<\left<\hat{C}_{st}^{}({\bf q},
\hat{\bf\Omega},\hat{\bf\Omega}^{\prime})\right>\right>_{f_s^{(N)},f_t^{(N)}}.
\nonumber\\
\end{eqnarray}
Here $\hat{C}_{st}^{}$ are Fourier transforms of the direct correlation
function. In practice we proceed as follows. Instability is signaled by
one of the eigenvalues of ${\cal T}$ or, equivalently,
the determinant of ${\cal T}$, becoming negative. 
Then, at fixed composition $x$, 
we look for the value of $\rho$ for which the four equations
\begin{eqnarray}
\hbox{det}\hspace{0.02cm}{\cal T}=0,\hspace{0.4cm}\frac{\partial}
{\partial q_{\mu}}\hbox{det}\hspace{0.02cm}{\cal T} =0,\hspace{0.4cm}
\mu=1,2,3
\end{eqnarray}
are satisfied simultaneously ($q_{\mu}=q_x, q_y, q_z$ for $\mu=1,2,3$, respectively).
In the case of the HSPC model, the C phase consists of
columns of particles arranged into a two-dimensional triangular lattice
perpendicular to the nematic director, and the two-dimensional ($xy$ plane)
perturbing wave that makes the fluid unstable will have a ${\bf q}$ vector
with non-zero $x$ and $y$ components. An instability associated with wave vectors
${\bf q}=(0,0,q)$, i.e., a wave along the $z$ direction (parallel to the
nematic director), will signal instability against smectic fluctuations.
For the system of HP the underlying lattice of the C phase is square,
so that instability will be signaled by 
wave vectors ${\bf q}=(q,0,0)$ or $(0,q,0)$.

\begin{figure}
{\centering \resizebox*{8cm}{!}{\includegraphics{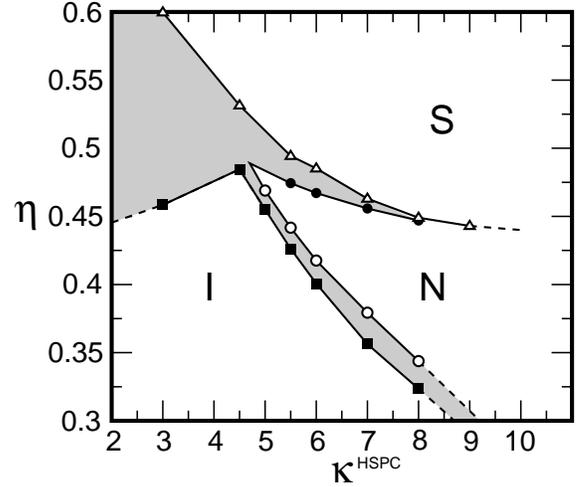}} \par}
\caption{\small Phase diagram of a pure system of
hard spherocylinders in the packing fraction ($\eta$)--aspect ratio ($\kappa
^{\rm HSPC}$) plane, as obtained from the EOT approach.
Labels are N, nematic; S, smectic; and I, isotropic.
}
\label{fig1}
\end{figure}

In the case of the FMT approximation the problem can be cast in matrix form
by introducing the fluctuation
\begin{eqnarray}
\delta\rho_s({\bf r},\hat{\bf\Omega})=\rho_s\sum_{\mu=1}^3
\epsilon_{s\mu}e^{i{\bf q}\cdot{\bf r}}\delta\left(\hat{\bf\Omega}-\hat{\bf e}_{\mu}\right).
\end{eqnarray}
We remark that in this case the decoupling approximation is not invoked, so the coupling between
positional and orientational degrees of freedom, assumed in the implementation
of the theory, is maintained. The parameters 
$\{\epsilon_{s\mu}\}$
are taken to be arbitrary, as it corresponds to a general fluctuation. In particular,
they are all taken to be independent, which is equivalent to saying that the values of the
order parameters $\{Q_s\}$ are allowed to vary as the perturbation is applied, in contrast
with the method used in the Onsager model. This is done for the sake of convenience, and
has no practical importance given that the condition for instability is searched for in the
space $(\rho,{\bf q})$, and the order parameters depend on the mean density $\rho$.
The ${\cal T}$ matrix now has six dimensions:
\begin{eqnarray}
{\cal T}_{s\mu,t\nu}=\frac{\delta_{st}\delta_{\mu\nu}}{\rho_{s\mu}}-
\hat{C}_{s\mu,t\nu}({\bf q}),
\end{eqnarray}
with
\begin{eqnarray}
C_{s\mu,t\nu}({\bf r}-{\bf r}')=
-\frac{\delta^2 \left[\beta {\cal F}_{\rm ex}\right]}{\delta \rho_{s\mu}({\bf r})
\delta\rho_{t\nu}({\bf r}')}\Bigg|_{\rho_{s\mu}({\bf r})=\rho_{s\mu}}
\end{eqnarray}
being the relevant direct correlation function and $\hat{C}_{s\mu,t\nu}({\bf q})$ its Fourier transform. Explicit expressions for this function will be presented
in the Appendix.

\begin{figure}
{\centering \resizebox*{8cm}{!}{\includegraphics{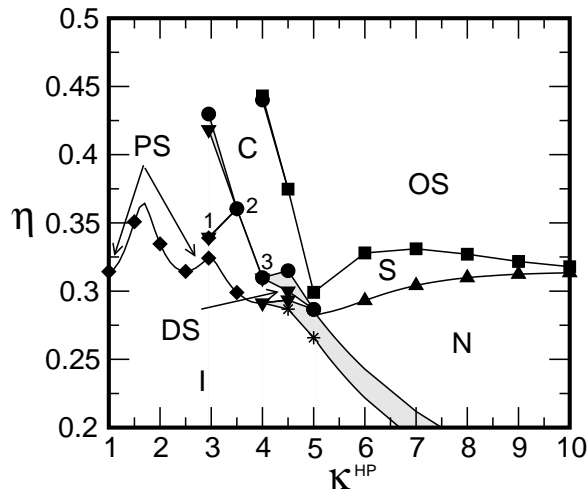}} \par}
\caption{\small Phase diagram in the packing fraction ($\eta$)--aspect ratio ($\kappa$)
for a pure component system of hard parallelepipeds, as obtained from the
FMT approach. Labels are N, nematic; S, smectic; 
OS, orientationally ordered solid; C, columnar; PS, plastic solid; and DS, discotic smectic.
The symbols indicate the points corresponding to calculated 
phase coexistence. The numbers indicate that the phase
transitions are of first order but that the scale is
too small to be seen by the eye.}
\label{fig2}
\end{figure}

\begin{figure}[h]
{\centering \resizebox*{8cm}{!}{\includegraphics{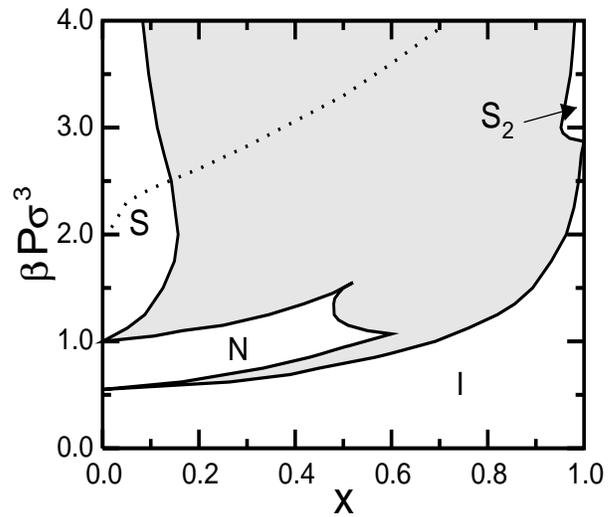}} \par}
\caption{\small 
Phase diagram in the pressure $P-$composition $x$ plane for a
mixture of HSPC with the same diameter $\sigma$ and aspect ratios
$\kappa_1=4.5$ and $\kappa_2=8.0$, as obtained from the EOT approach.
Reduced units are used for length ($\sigma$) and energy
($kT$, the thermal energy). The continuous lines indicate first-order phase transitions.
The shaded regions are the two-phase regions of phase coexistence. The regions of stability
are labeled by S (standard smectic formed by layers identical in composition), 
N (nematic), I (isotropic), and S$_2$
(microsegregated smectic phase with long particles located in the interlayer
space). The dotted line
is the spinodal line corresponding to the instability of the nematic phase with respect
to columnar-type fluctuations.}
\label{fig3}
\end{figure}

\begin{figure}[h]
{\centering \resizebox*{8cm}{!}{\includegraphics{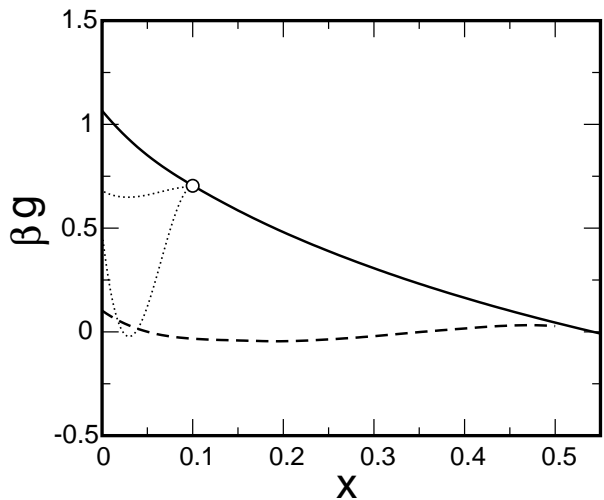}} \par}
\caption{\small 
Gibbs free energy per unit volume and unit thermal energy $\beta g$
as a function of composition $x$ for the nematic phase (continuous line)
and the smectic phase (dashed line) for a mixture of HSPC with aspect ratios
$(L_1+\sigma)/\sigma=4.5$ and $(L_2+\sigma)/\sigma=8.0$. A linear term
$34.35-12.52x$ has been subtracted so as to better see 
the curvature of the curves.
The circle is the bifurcation point where the nematic phase becomes metastable
with respect to columnar-type fluctuations. The dotted lines 
are two possible
free-energy branches for the columnar phase that could result from a
full minimization of the free-energy functional.} 
\label{fig3bis}
\end{figure}

\begin{figure}[h]
{\centering \resizebox*{8cm}{!}{\includegraphics{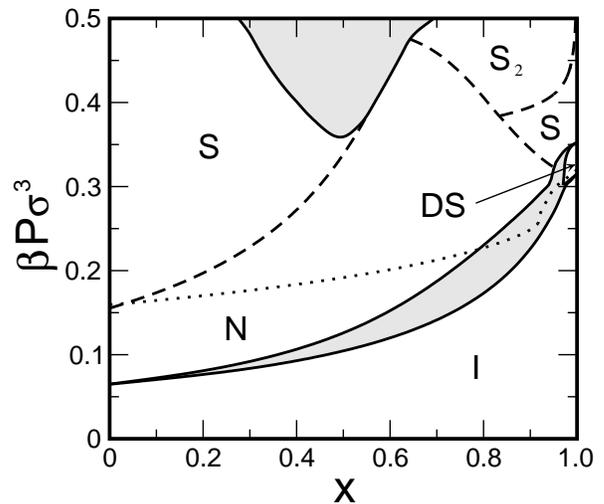}} \par}
\caption{\small Phase diagram of the HP mixture in the FMT approach,
represented in the pressure-composition plane (reduced units are
used for the pressure). All particles have the same cross section $\sigma^2$, 
and its aspect ratios are $\kappa_1=4.5$ and $\kappa_2=8$. 
The continuous lines indicate the first-order phase transitions.
The shaded areas are the regions of two-phase coexistence. The dashed lines 
indicate the continuous transitions between the nematic (N) and smectic (S) phases, 
or between different S phases.
The dotted line is the spinodal line signaling instability of the N phase
against spatial fluctuations (abrupt change in slope is associated with
rapid variation of nematic order parameter).
%plastic solid (when the parent phase is isotropic) phases.
S phase is a standard smectic phase. S$_2$ phase consists of
alternating layers of different composition. 
DS is a discotic smectic phase.
}
\label{fig4}
\end{figure}

\begin{figure}
{\centering \resizebox*{8cm}{!}{\includegraphics{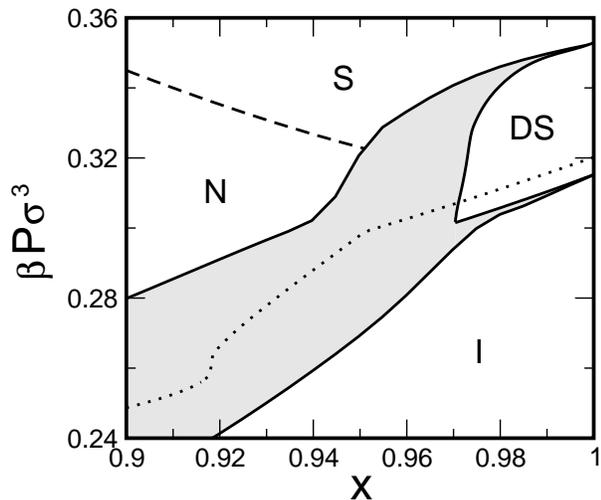}} \par}
\caption{\small A zoom of a particular region of the phase diagram represented in Fig. \ref{fig4}
around the DS phase. For a key to lines and labels see caption of Fig. \ref{fig4}.}
\label{fig5}
\end{figure}

\begin{figure}
\mbox{\includegraphics*[width=7cm,angle=0]{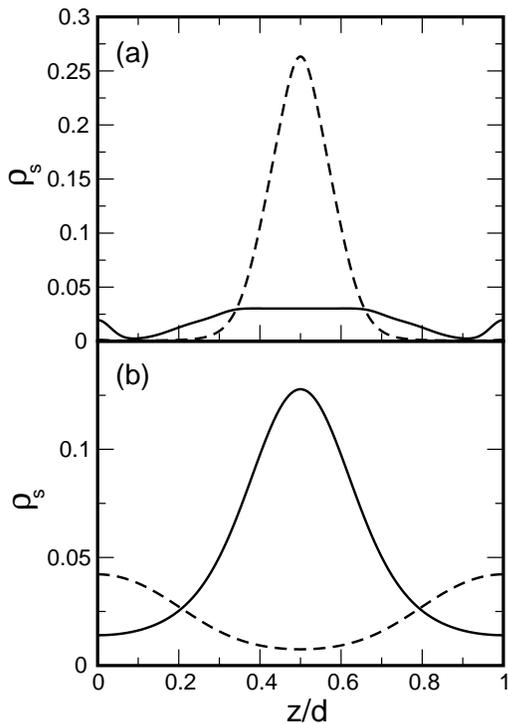}}
\caption{\small Density profiles of short (solid line) and long
(dashed line) particles of coexisting (a) S and 
(b) S$_2$ phases at pressure $\beta P\sigma^3=0.5$. 
The values of coexistence parameters are $x=0.276$, $\eta=0.4768$, and $d/L_2=1.1642$
for S phase, and $x=0.693$, $\eta=0.4185$, and $d/L_1=1.1498$ for S$_2$ phase.
}
\label{fig6}
\end{figure}

\section{Results}
\label{Results}

Although the HSP and HP models are geometrically different (HSP particles
have a semispherical cap which is absent in HP particles, and their
sections are circular and square, respectively), and their theoretical 
treatment is quite different, the corresponding phase diagrams will
be compared in this section, in the hope that we can extract some
useful information about their actual phase behavior. This comparison 
can be made provided we use comparable aspect ratios
$\kappa_s$; we should take into account the relation
\begin{eqnarray}
\kappa_s^{\hbox{\tiny HSPC}}\equiv
\left(\frac{L_s+\sigma}{\sigma}\right)_{\hbox{\tiny HSPC}}
\longleftrightarrow
\kappa_s^{\hbox{\tiny HP}}\equiv\left(\frac{L_s}{\sigma}\right)_{\hbox{\tiny HP}}.
\label{10}
\end{eqnarray}

\subsection{Phase diagrams for pure components}

Before showing the results for mixtures, and
in order to have a feeling as to how the two theoretical models can be
compared and how definite predictions can be drawn from this comparison,
it is convenient to recall the results obtained for the corresponding
one-component systems using the same theoretical models
that will be used later for mixtures. The phase diagrams
for the pure systems are shown in Figs. \ref{fig1} and \ref{fig2}, both
in the packing fraction-aspect ratio plane (we should 
remind the definitions of the corresponding aspect
ratios, $\kappa^{\rm HSPC}$ and $\kappa^{\rm HP}$,
respectively). A quick comparison of the two
phase diagrams indicates that the IN phase transition
is qualitatively similar: it is of first order and, for decreasing
aspect ratio, ends at some value of $\kappa$, below which the N phase
ceases to be stable and is superseded by the spatially
ordered S phase or other more ordered phases.
For HP particles the phase diagram has been analyzed in some
detail using the FMT approach\cite{Yuri1} 
(see Fig. \ref{fig2}); it is relatively complex, with
the C phase becoming stable in a narrow range
of aspect ratios, a discotic smectic (DS) phase and
also a plastic solid (PS). At high packing fractions 
and $\kappa^{\rm HP}\agt 5$ the stable phase is an
orientationally ordered solid (OS). The prediction of a DS phase
has been confirmed by simulations of this particular model. \cite{Harrowell}
The phase diagram for HSPC obtained with the EOT has not been 
analyzed in such  great detail (see Fig. \ref{fig1}). It contains an IN
transition ending at some value of $\kappa$ which is very close
to that predicted by the FMT approach for HP particles.
The NS transition shows a negative slope, compared with the positive
slope obtained in the FMT (at least in the range $5<\kappa<10$), 
and it is of first order up to some value of $\kappa$, above which it
becomes second order. By contrast, the NS transition in the FMT approach
is always of second order. Since neither the crystalline nor the plastic-solid
phases have been explored in the EOT model (mainly due to the
fact that this model is not expected, by construction, to provide good 
results for phases with full spatial order) the comparison of the
phase diagrams cannot be extended beyond the liquid-crystalline phases.
This is not a limitation for our present purposes, since we would like
to focus our investigation on these phases.
However, simulations on HSPC indicate that at high packing fractions 
there appears a solid phase with full order, which becomes a plastic
solid when the aspect ratio is low. A more elaborate theory (such as
that due to Somoza and Tarazona\cite{Somoza}) should be used to
address this question.

\subsection{Phase diagrams for mixtures}

We have first investigated the phase behavior of a mixture of particles
with $(\kappa_1,\kappa_2)=(4.5, 8.0)$ and $\sigma_2/\sigma_1=1$.
%and $(\kappa_1,\kappa_2)=(6.0, 8.0)$.
The results are contained in Figs. \ref{fig3}--\ref{fig5}. 
Throughout this section phase diagrams will be presented in the 
pressure $p$--composition $x$ plane (by convention,
we take the composition of the mixture to be given by the
variable $x\equiv x_1$, where species 1 is chosen to correspond to the
shortest particles). Figure \ref{fig3} shows the results as
obtained from the Onsager theory (this phase diagram has already been
presented in Ref. 4). The most representative feature
of this phase diagram is the strong segregation of the S phase,
which largely preempts the I-N transition.
Two S phases appear: the standard S phase, with identical
smectic layers, and the S$_2$ phase, where layers of different
compositions alternate: this is a microsegregated phase.
This phase is characterized by the two
variational parameters $\{\lambda_s\}$ having different signs,
which implies the density distributions of the two species
being shifted one with respect to the other by half a smectic
period. Below a reduced pressure $\beta P\sigma^3=2.9$ there is direct
coexistence between the S and I phases, which ends in a S-N-I triple
point. Above this pressure the S phase coexists with the S$_2$
phase. The Zwanzig model for a comparable mixture is shown in Figs. 
\ref{fig4} and \ref{fig5}.
Note that the pressure scale is very much reduced with respect to
that from the Onsager theory, so that a good reference to compare
the two phase diagrams is the location of the I-N transition, which
is qualitatively similar to the previous case. In the Zwanzig model
S segregation appears at a much higher pressure, and does not
preempt the I-N transition. There is no direct S-I transition, which
is superseded by continuous S-N transitions, and at high pressure there
appears a first-order transition between the standard S phase and
the two-layer smectic phase S$_2$, the latter appearing when the
short-particle component is more abundant; this is in agreement with
the predictions based on the Onsager model. However, an
important difference is that the S and S$_2$ phases undergo
continuous phase transitions to the N phase, a feature that
is absent in the Onsager model. This is to be expected since 
in the one-component HP fluid the N-S transition is always of
second order. Also, in a model with restricted orientations the S 
phase is largely destabilized with respect to the N phase, so that
the region where the N phase is stable is considerably larger 
in the Zwanzig model than predicted by the EOT. 
In addition, the FMT approach predicts also
a continuous transition between the S and S$_2$ phases in the region
where most of the particles correspond to the shorter component.
This feature is absent in the EOT approximation. 

\begin{figure}[h]
\mbox{\includegraphics*[width=7cm,angle=0]{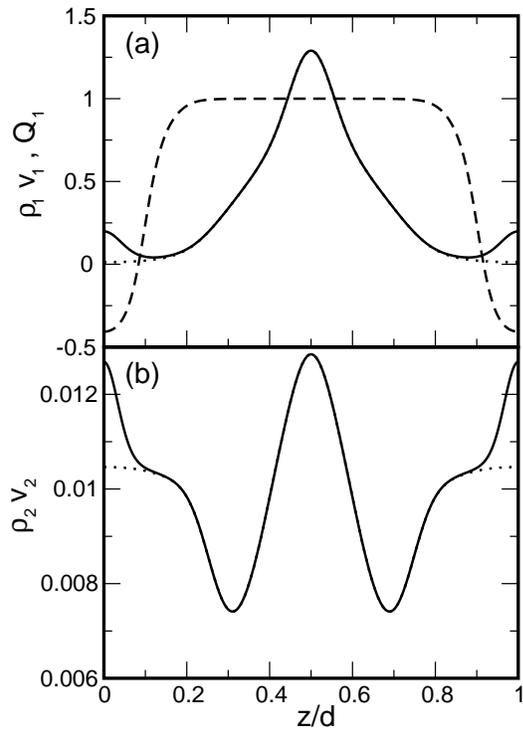}}
\caption{\small Density (solid line) and order-parameter (dashed 
line) profiles of the species with (a) short and (b) long particles.
Densities are scaled with the corresponding particle volume $v_s$.
Density profiles of each species 
along a direction parallel to the director are plotted with dotted lines. 
This smectic phase corresponds to a state point in the phase diagram with
$\beta P\sigma^3=0.442$ and $x=0.9858$.
The smectic period in units of the small species length is 
$d/L_1=1.3717$, and the mean total packing fraction $\eta=0.3934$.}  
\label{fig7}
\end{figure}

\begin{figure}[h]
\mbox{\includegraphics*[width=7cm,angle=0]{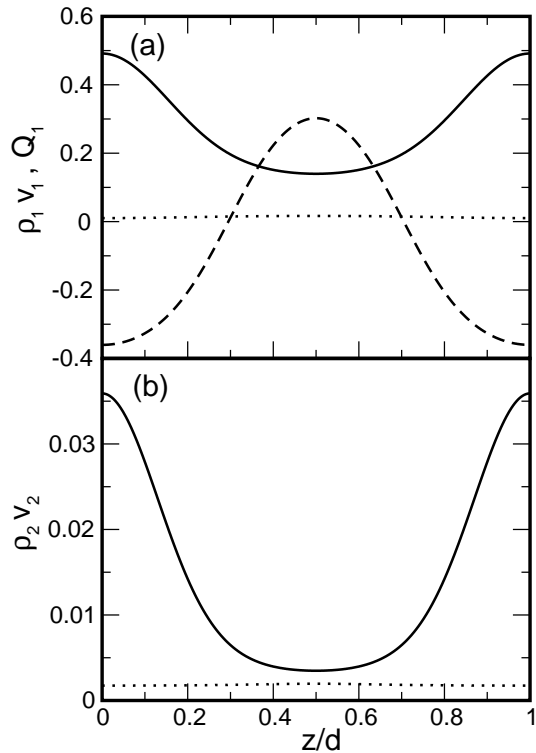}}
\caption{\small Density (solid line) and order-parameter (dashed line) profiles 
corresponding to the species with (a) short and (b) long particles for the
discotic-smectic phase.
Densities are scaled with the corresponding particle volume $v_s$.
The projection of the density distribution of each species along
the direction of the director is plotted using dotted lines
(note that these distributions have maxima at $z/d=0.5$).
The discotic smectic phase corresponds to the I-N-DS triple point 
with $\beta P\sigma^3=0.302$ and $x=0.9715$. The smectic 
period in units of the cross-section length is $d/\sigma=1.2850$, and 
the mean total packing fraction $\eta=0.2915$.}
\label{fig8}
\end{figure}

In order to understand the structure of the S phases in more detail,
it is interesting to examine the density and order-parameter profiles
along one smectic period $d$ (some of these profiles were shown in Ref.
4 for the EOT model, so here we will only show the profiles 
corresponding to the FMT approach). This is done in Figs. \ref{fig6}--\ref{fig8}.
In Fig. \ref{fig6} the profiles at a state point with reduced pressure
$\beta P\sigma^3=0.5$ on the S-S$_2$ coexistence line are shown.
As can be seen the S phase [Fig. \ref{fig6}(a)] is composed of identical
layers, with maxima in the density distribution of both species
coinciding at the center of each layer. An interesting point to mention 
is that the position of the short particles 
is delocalized over the length of the long particles 
as can be seen from Fig. \ref{fig6}(a). 
By contrast, in the S$_2$
phase, these maxima are shifted by half a period (an amount equal to $d/2$), 
so that the layers with different compositions alternate. Figure \ref{fig7}
shows profiles at some particular state point on the S-S$_2$ second-order transition
line. In this point most of the particles in the system are short, and the 
few long particles present are evenly distributed among the layers and the
interstitials. This corresponds to the second-Fourier coefficient 
$\alpha^{(1)}_{21}$ of the
density distribution of the long particles becoming zero at this state point.
As $x$ is reduced long particles tend to predominantly populate the insterstitials,
defining a bilayer structure. On the contrary, as $x$ is increased, the 
long particles tend to arrange into the layers formed by the particles of
the other component. This behavior is at variance with the predictions of
the EOT approach, which imply that long particles added to a smectic phase
made up of the short particles tend to populate the interstitials, regardless
of their concentration. In Fig. \ref{fig7} the density profiles of each species 
along a direction parallel to the director are also plotted with dotted lines. 
These profiles indicate that particles in the layers are mostly directed
along the layer normal, whereas short particles tend to align in-plane in the 
interstitials, and a small amount of long particles also adopt this orientation. 
As can readily be seen, the order-parameter profile also supports this conclusion.

An interesting feature of the phase diagram in Fig. \ref{fig4} is that,
for a large concentration of short particles, there is a small region 
where a discotic-smectic (DS) phase
is stable (see Fig. \ref{fig5}, which is an enlargement of Fig. \ref{fig4}
in the region where the DS phase appears). 
The DS phase was found in a one-component system of HP (Ref. 11)
(see Fig. \ref{fig2}, which indicates that for a one-component system with 
$\kappa=4.5$ there appears a DS phase at packing fractions $\eta\approx 0.3$)
and confirmed by simulation. \cite{Harrowell}
The structure of the DS phase can be understood by examining Fig. \ref{fig8}.
We see that the phase consists of a succession of identical layers but,
in contrast with the usual smectic phases, particles in the layers are
almost exclusively oriented in plane, with nematic order parameter 
large and negative. The interstitials contain a considerable amount
of particles, slightly oriented along the layer normal. The long-particle
component tends to follow the majority component and populate the layers. 
The present result for the mixture indicates that as soon as
a small amount of long rods is added to the pure fluid, the DS phase
becomes very rapidly unstable.

\begin{figure}[h]
{\centering \resizebox*{8cm}{!}{\includegraphics{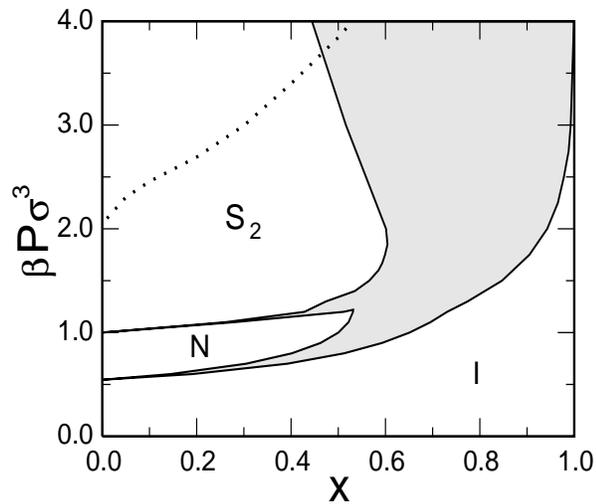}} \par}
\caption{\small 
Phase diagram in the pressure $P-$composition $x$ plane for a
mixture of HSPC with the same diameter $\sigma$ and aspect
ratios $\kappa_1=6.0$ and $\kappa_2=8.0$, as obtained from the EOT approach. 
See caption of Fig. 1 for key to labels and lines.}
\label{fig9}
\end{figure}

\begin{figure}[h]
\mbox{\includegraphics*[width=8cm,angle=0]{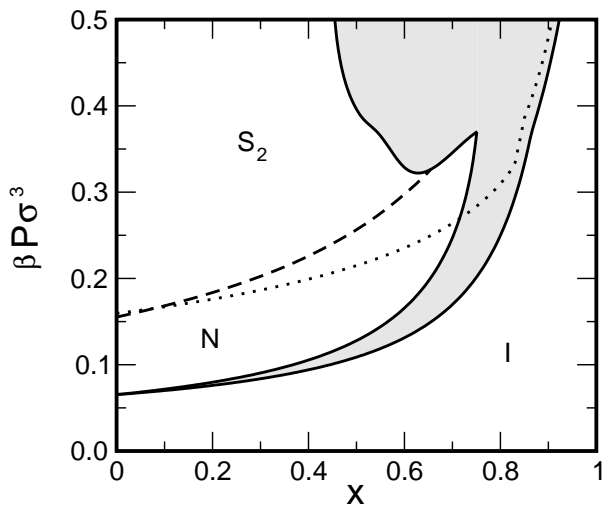}}
\caption{\small Phase diagram of a binary mixture of hard parallelepipeds
($\kappa_1=1$) and cubes ($\kappa_2=8$) as obtained from the FMT approach. 
See caption of Fig. \ref{fig4}
for key to labels and lines.}
\label{fig10}
\end{figure}

\begin{figure}
\mbox{\includegraphics*[width=7cm,angle=0]{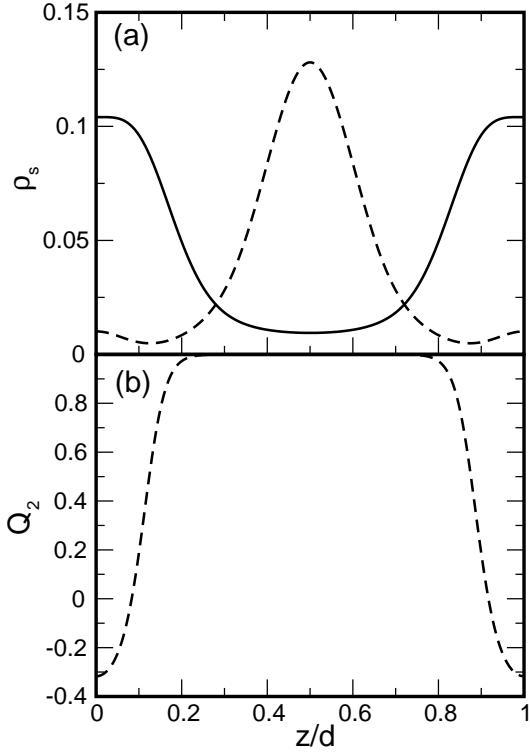}}
\caption{\small (a) Density profiles of cubes (solid line) and rods 
(dashed line) for the coexisting S$_2$ phase ($\beta P\sigma^3=
0.37$, $x=0.543$, $\eta=0.3579$, and $d/L_2=1.6422$)
at the I-N-S$_2$ triple point shown in Fig. \ref{fig10}.
(b) Order-parameter profile of rods for the same thermodynamic state.} 
\label{fig11}
\end{figure}

\begin{figure}[h]
\mbox{\includegraphics*[width=7cm,angle=0]{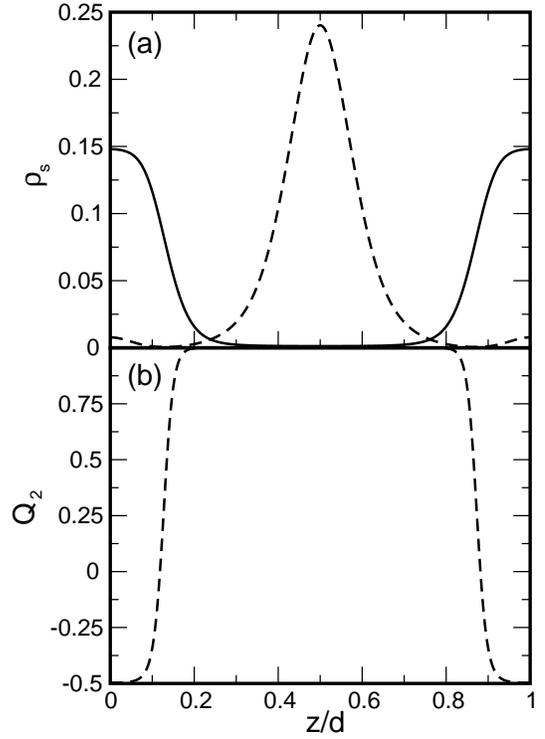}}
\caption{\small (a) Density profiles of cubes (solid line) and rods
(dashed line) of the S$_2$ phase which coexists with the I phase
at $\beta P\sigma^3=0.5$, $x=0.455$, $\eta=0.4414$, and $d/L_2=1.5139$
(see Fig. \ref{fig10}).
(b) Order-parameter profile of rods for the same thermodynamic state.}
\label{fig12}
\end{figure}

As an example of another mixture showing strong smectic segregation,
but with one species composed on nonmesogenic particles, we discuss
the cases $\kappa_1=1$, $\kappa_2=8$, and $\sigma_2/\sigma_1=1$, i.e., mixtures of HSPC with 
hard spheres or HP with hard cubes (see Figs. \ref{fig9} and \ref{fig10}). 
Here again the two models give the same phase
behavior: at high volume fractions of the nonmesogenic component we
find an isotropic phase, and a relatively small addition of this
component to a fluid composed mainly of long particles destabilizes
both liquid-crystalline phases. Again the N-S phase is of first order
in the Onsager model but of second order in the Zwanzig theory. Apart
from that, the two phase diagrams are qualitatively similar. 
The structure of the smectic 
phase is interesting, as it is again a clear example of a microsegregated
phase: layers rich in long particles alternate with layers made up
of (almost exclusively) hard spheres (or hard cubes, as the case may be). Figures 
\ref{fig11} and \ref{fig12}, containing density profiles, show
this feature for the Zwanzig model (in the case of the EOT
the segregation into well-defined layers is even more pronounced 
\cite{Cinacchi1}).
Both models predict that the smectic phase is destabilized with respect to the
nematic phase as the nonmesogenic component is added\cite{depletion}
(the slope of the 
N--S$_2$ phase boundary in the pressure--composition plane is positive).

\begin{figure}
\mbox{\includegraphics*[width=8cm,angle=0]{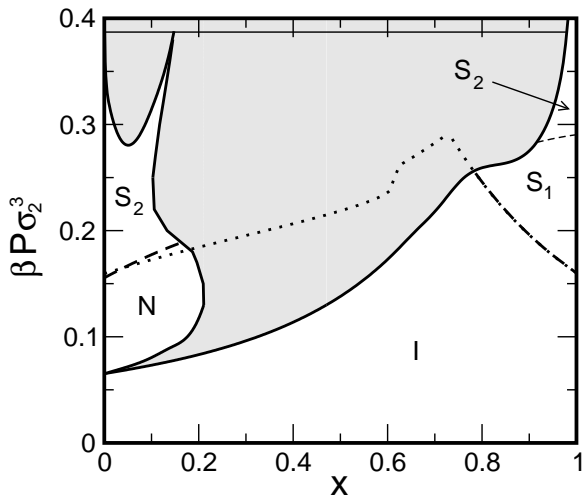}}
\caption{\small Phase diagram of a binary mixture of cubes, $\kappa_1=1$,
and rods, $\kappa_2=8$, with $\sigma_1/\sigma_2=2$ (i.e., with equal 
particle volumes), as obtained from the FMT approach.
The pressure at which the three smectic phases coexist at a triple point 
is indicated by a horizontal line. See caption of Fig. \ref{fig4} for key to
labels and lines.
}
\label{fig13}
\end{figure}

\begin{figure}[h]
\mbox{\includegraphics*[width=7cm,angle=0]{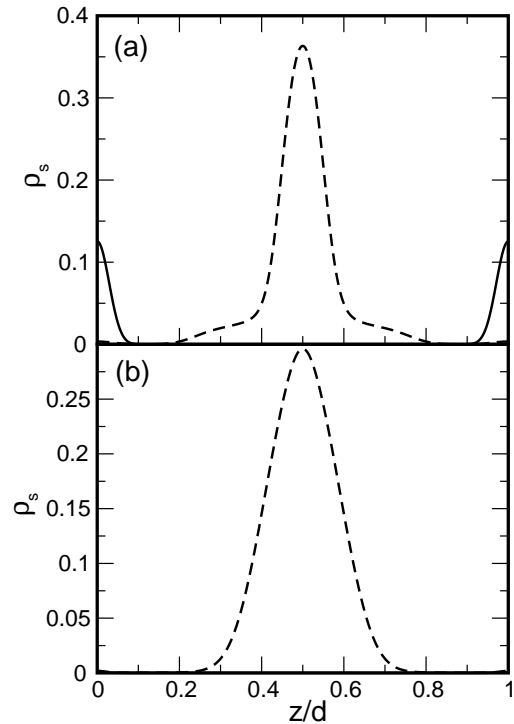}}
\caption{\small Density profiles for two of the S$_2$ phases that
coexist at the triple point with $\beta P\sigma_2^3\approx 0.388$
(see Fig. \ref{fig13}).
(a) Density profiles of cubes (solid line) and
rods (dashed line) corresponding to the coexisting S$_2$ phase
with $x\approx 0.148$ ($\eta \approx 0.4816$, $d/L_1\approx 1.4384$). 
(b) Density profile for the coexisting S$_2$ phase
with $x_1\approx 0.0007$ ($\eta\approx 0.4902$, $d/L_1\approx 1.1934$). 
}
\label{fig14}
\end{figure}

\begin{figure}[h]
\mbox{\includegraphics*[width=7cm,angle=0]{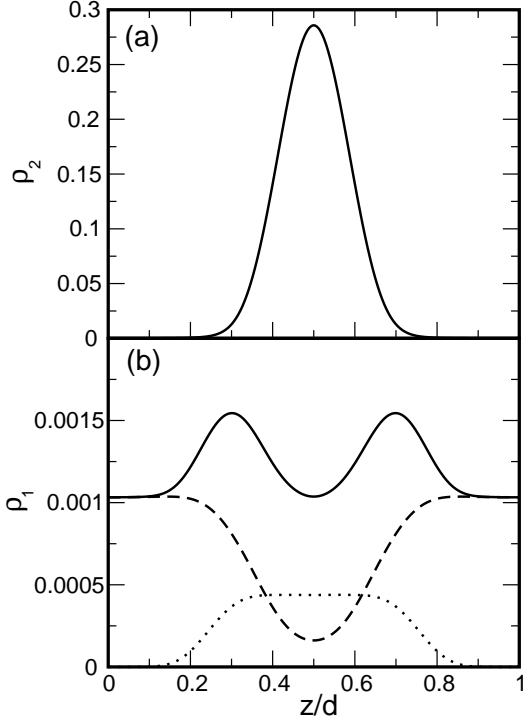}}
\caption{\small (a) Density profile of cubes corresponding 
to the coexisting S$_2$ phase at the triple point of Fig. \ref{fig13}
with composition $x\approx 0.98$ ($\eta \approx 0.4860$, $d/L_2\approx
1.1839$). (b) Density profile of rods at the same thermodynamic
conditions; the dotted line represents the density profile 
corresponding to the rods that lie perpendicular to the
director, whereas the dashed line refers to the particles lying 
parallel to the director. The total density profile is
represented by the solid line.}
\label{fig15}
\end{figure}

As a final mixture we have studied, using only the Zwanzig approximation,
the case where the parallelepipeds and the cubes have the same 
volumes but different breadths, i.e., different cross sections. 
The phase diagram corresponding to the case $L_1/\sigma_1=8$, $L_2/\sigma_2=1$, 
$\sigma_2/\sigma_1=2$ is shown in Fig. \ref{fig13}. Due to the different
cross-sectional areas of the particles the smectic phases strongly
segregate at high pressures. The IN segregation region is also
greatly enhanced. A remarkable feature of the phase diagram is the
existence of demixing in the region where the S$_2$ phase is stable at low 
compositions; this
phase transition ends in a lower critical point and, at some higher
pressure, transforms into a triple point where the two bilayer smectic phases
with moderate content of cubes coexist with a third one which is mainly
formed by cubes. The density profiles in Figs. \ref{fig14} and \ref{fig15}
are useful to understand the structure of these phases. In Fig. \ref{fig14}
the density distribution of the first two S$_2$ phases is shown. Figure \ref{fig14}(a)
shows that there is a very strong microphase separation between the two
components which therefore arrange in layers of almost pure composition. 
The origin of this isostructural phase separation into two bilayer structures
can be understood in terms of the so-called depletion effect, which has
an entropic origin: the cubes added to a pure mixture of rods in a smectic
configuration arrange themselves in the interstitials between the smectic 
layers, and this arrangement creates an effective attraction between the
layers, giving rise to a first-order ``condensation'' phase transition.

The structure of the bilayer smectic phase at high pressure and with a
high content of cubes is also very interesting as it illustrates another effect
of the depletion interactions; this is shown in Fig. \ref{fig15}.
The thermodynamic state corresponds to the phase that coexists at the triple 
point, with $x\approx 0.98$. From Fig. \ref{fig15}(a) we can see that the cubes
arrange themselves into well-defined layers; however, the rods adopt a very
different configuration since they do not occupy the instertitials between 
layers in a uniform manner: an excess density of rods develops right at
the edges of the main distribution of cubes.
This is represented in Fig. \ref{fig15}(b) by the solid line,
which in turn has been split into parallel and perpendicular components
to the director (layer normal) to show that the rods in the excess regions
are oriented approximately with equal probability with respect to the 
three mutual perpendicular directions. We can conclude that 
the cubes act as a
soft wall against which rods are piledup as if adsorbed.


\subsection{Spinodal lines for columnar order}

The discussion thus far has been focused on the phase diagrams containing 
I, N, and
S phases. As discussed in the introduction, there are strong arguments that
suggest that the columnar phase might play a role in the phase stability
of the mixtures of hard rods. Using the methodology outlined in Sec. 
\ref{columnar}, we have calculated the spinodal line where the nematic phase 
becomes unstable with respect to fluctuations involving local spatial order.
As implicit in the presentation of the method, our approach does restrict the 
search for instability to any particular structure; usually, however, the 
first instability (at the lowest mean density) is found for fluctuations against 
columnar-type fluctuations (this is signaled by a wave vector with nonvanishing 
$x$ or $y$ component), except in a few cases to be discussed below.

The spinodal line is represented in the phase diagram of Fig. \ref{fig3},
corresponding to the mixture with $\kappa_1=4.5$, and $\kappa_2=8$, 
$\sigma_2/\sigma_1=1$, as analyzed in the context of the EOT approach.
As can be seen from the figure, the instability occurs
at relatively high pressure, which means that there is a relatively
wide region of stability for the smectic phase and consequently, the predictions
on smectic demixing in mixtures of HSPC are 
plausible. However, our results indicate that the presence of
smectic segregation should be taken with some caution, since it could be 
preempted by direct coexistence between the C and S$_2$ phases. Without a
full calculation of the free energy of the C phase this question cannot
be settled completely. In order to discuss this point, we plot in Fig. 
\ref{fig3bis} the free-energy branches $g(x)$ of the N and S phases for a
mixture with fixed pressure $\beta P\sigma^2=2.43$.
The bifurcation point is indicated by a 
circle. Starting at the bifurcation point at $x=0.1$ two possible free-energy branches for the C 
phase have been represented (the C-phase branch will exist for values 
$x\le 0.1$, see Fig. \ref{fig3}). One possibility is that the curve 
does not cross the smectic-phase branch: in this case the C phase exists
as a metastable phase and the S phase is always more stable. The alternative
situation is the one corresponding to the other curve, which crosses 
the smectic
branch: now the C phase becomes more stable and coexists with
the S phase. This demonstrates that the bifurcation analysis can only
provide trends as to the possible stabilization of the phases and that,
once a bifurcation point is shown to exist, one cannot rule out any 
scenario.

The FMT theory on the HP mixture shown in 
Fig. \ref{fig4} predicts a spinodal line at low pressure,
below the corresponding line associated with smectic ordering (there
is a small region, for small values of $x$, where the smectic line
is slightly above the columnar line which leaves a small region of
smectic stability; see Figs. \ref{fig4} and \ref{fig10}). Clearly, 
the rich smectic phase behavior predicted by the theory is completely
preempted by columnar ordering.
%In Fig. \ref{fig3} it can be seen that the spinodal line is not
%smooth but shows an abrupt change (indicated by a vertical arrow). 
%This is due to the fact that the perturbed phase is an isotropic phase, 
%instead of the nematic phase. 
At high composition $x$ the spinodal line ends
in the region of DS stability; this means that the DS phase preempts
the instability of the I phase against columnar or plastic-solid fluctuations.

Now we come to show the results for the mixture with $\kappa_1=1$, 
$\kappa_2=8$, and $\sigma_2/\sigma_1=1$. The results from the EOT for
the HSPC model are shown in Fig. \ref{fig10}. We can see that the 
spinodal line shows the same trend: it increases with composition,
and is located at relatively high pressure so that a wide region
of smectic stability is found. By contrast, the FMT approach 
(Fig. \ref{fig11}) again predicts a spinodal line against columnar-type
fluctuations below the region of smectic stability, save for a small
interval in composition at low values of $x$.

Finally, we have also calculated the spinodal line associated with columnar
fluctuations for the mixture with equal particle volumes but different
cross sections,
$L_1/\sigma_1=8$, $L_2/\sigma_2=1$, and $\sigma_2/\sigma_1=2$. The results
are in the same line as in the previous cases: the smectic phases are
completely preempted by the instability against columnar fluctuations.
Note that the slope of the spinodal line changes sign at $x\approx 0.7$
and then becomes a spinodal line for instability of the I phase
($0.75\alt x\le 1$) against spatial fluctuations; in this case
the divergence of the structure factor occurs at the same density $\rho$,
independent of the orientation of the wave vector, ${\bf q}=(0,0,q)$,
$(0,q,0)$, or $(q,0,0)$, which means that the instability could in 
principle be associated with the appearance of columnar, smectic, or
solid order. This result is analogous to the behavior observed in 
one-component hard cubes \cite{Yuri1} for which it turns out that 
the smectic, columnar, and solid free-energy branches start at the 
same bifurcation point, the solid phase being the most stable phase. Only 
a complete density-functional minimization could elucidate the relative 
stability between all these phases in the mixture.  

\section{Conclusions}
\label{Conclusions}
In this paper we have analyzed the phase behavior of binary 
mixtures of hard particles
having different geometries, with a view to locating the conditions under
which spatially ordered phases are formed. Two different theoretical,
somehow complementary, approaches have been used: one based on the standard 
Onsager theory for hard particles, the other being a fundamental-measure theory 
for hard parallelepipeds in the Zwanzig approximation, i.e. with restricted
particle orientations. It is known that the tendency of the two species 
to macroscopically 
segregate is enhanced when the aspect ratio of the particles
is more dissimilar. For the more extreme cases this creates large
demixing regions where smectic phases of different compositions coexist,
even preempting part of the isotropic-nematic transition. 
Even more significant, in some mixtures the smectic phase exhibits
microphase segregation, where the species demix at the level of the
one-dimensional smectic unit cell. Microphases may undergo transitions
to smectic phases with the usual homogeneous layer composition. 

We have demonstrated that these mixtures tend to form columnar phases,
even when the cross-sectional length of the particles is unequal, regardless
of their relative length. The extent to which columnar order is more
or less stable than smectic order has been analyzed using bifurcation theory.
In general, the bifurcation analysis of the EOT model predicts that columnar 
order may appear; whether or not this order preempts direct smectic-phase 
segregation cannot be ascertained with the present methodology. However,
it is clear that large regions of smectic stability still persist in the 
phase diagram. By contrast, the FMT approach in the
Zwanzig approximation predicts that smectic order is almost completely 
suppressed. The difference in behavior predicted can be traced back
to the different treatments of orientational fluctuations (in the latter
case orientations are restricted to lie along three orthogonal directions,
which enhances columnar order since the order parameter is usually very high). 
The results obtained with the
two approaches probably set the limits of what should be expected in the
real mixtures. In this respect, given the lack of more reliable 
theoretical treatments that can give more quantitative answers and the 
tremendous difficulties involved in making theoretical progress, one
should rely on computer simulations, an approach that is feasible but
unfortunately not pursued so far.

\section*{ACKNOWLEDGMENTS}
One of the authors (Y. M.-R.) 
was supported by a Ram\'on y Cajal research contract from Ministerio 
de Educaci\'on y Ciencia (Spain). This work is part of research projects
Nos. BFM2003-0180, BFM2001-0224-C02-01,
BFM2001-0224-C02-02, and BFM2001-1679-C03-02 of the Ministerio de 
Educaci\'on y Ciencia (Spain).

\appendix*
\section{Direct correlation function in the FMT approach.}

In this section we present explicit expressions for the direct correlation
function of the mixtures. 
In the FMT approach we have, in the nematic phase,
\begin{eqnarray}
&-&C_{s\mu,t\nu}({\bf r}-{\bf r}')\nonumber \\&\equiv& 
\frac{\delta^2\left[\beta{\cal F}_{\rm ex}\right]}{\delta \rho_{s\mu}({\bf r})
\delta\rho_{t\nu}({\bf r}')}\Bigg|_{\rho_{s\mu}({\bf r})=\rho_{s\mu}=
\rho x_s\left[1+\left(3\delta_{\mu z}-1\right)Q_s\right]/3}\nonumber \\
&=&\frac{\langle \omega^{(0)}_{s\mu}\ast\omega^{(3)}_{t\nu}\rangle+
\langle \boldsymbol{\omega}_{s\mu}^{(1)}\ast 
\boldsymbol{\omega}_{t\nu}^{(2)}\rangle}{1-\xi_3}\nonumber\\&+&
\frac{\boldsymbol{\xi}_2\left[\langle \boldsymbol{\omega}_{s\mu}^{(1)}\ast 
\omega^{(3)}_{t\nu}\rangle+\boldsymbol{\Omega}^{(2)}_{s\mu,t\nu}\right]}
{(1-\xi_3)^2}\nonumber\\
&+&\left[\frac{\boldsymbol{\xi}_1}{(1-\xi_3)^2}+
\frac{2\boldsymbol{\zeta}_2}{(1-\xi_3)^3}\right]
\langle \boldsymbol{\omega}_{s\mu}^{(2)}\ast\omega_{t\nu}^{(3)}\rangle\nonumber\\&+&
\left[\frac{\xi_0}{(1-\xi_3)^2}+
\frac{2\boldsymbol{\xi}_1\boldsymbol{\xi}_2}{(1-\xi_3)^3}+
\frac{6\xi_{2x}\xi_{2y}\xi_{2z}}{(1-\xi_3)^4}\right]
\omega_{s\mu}^{(3)}\ast\omega_{t\nu}^{(3)},\nonumber\\
\label{lac}
\end{eqnarray}
The symbol $\ast$ stands for
convolution: $f\ast g=\int d{\bf r}'f({\bf r}')g({\bf r}-{\bf r}')$
and it is introduced as a shorthand notation.
All expressions inside the angular brackets are understood to be 
symmetrized with respect to their indices $s\mu$, $t\nu$. For example, 
\begin{eqnarray}
\langle \omega^{(0)}_{s\mu}\ast \omega^{(3)}_{t\nu}\rangle=
\omega^{(0)}_{s\mu}\ast \omega^{(3)}_{t\nu}+
\omega^{(0)}_{t\nu}\ast \omega^{(3)}_{s\mu}.
\end{eqnarray}
Also, all boldface variables are three-dimensional vectors 
\begin{eqnarray}
\boldsymbol{\xi}_{\alpha}=(\xi_{\alpha x},\xi_{\alpha y},\xi_{\alpha z}),\hspace{0.4cm}
\boldsymbol{\omega}^{(\alpha)}_{s\mu}=
\left(\omega_{s\mu}^{(\alpha x)},\omega_{s\mu}^{(\alpha y)}, 
\omega_{s\mu}^{(\alpha z)}\right),\nonumber\\
\end{eqnarray}
with $\alpha=1,2$
and the products between any two of them are to be taken as scalar products. 
The expression for $\boldsymbol{\Omega}^{(2)}_{s\mu,t\nu}$ is 
\begin{eqnarray}
\boldsymbol{\Omega}^{(2)}_{s\mu,t\nu}=
\left(\langle\omega_{s\mu}^{(2z)}\ast\omega_{t\nu}^{(2y)}\rangle,
\langle\omega_{s\mu}^{(2x)}\ast\omega_{t\nu}^{(2z)}\rangle,
\langle\omega_{s\mu}^{(2y)}\ast\omega_{t\nu}^{(2z)}\rangle\right).\nonumber\\
\hspace*{5.cm} (A.4) \nonumber
\end{eqnarray}
The expressions for $\xi_0$ and $\xi_3$ and for the components of
$\boldsymbol{\xi}_{\alpha}$ and $\boldsymbol{\zeta}_2$ are
\begin{eqnarray}
\xi_0&=&\rho,\quad \xi_3=\rho \sum_s x_sv_s,\nonumber\\
\xi_{1\beta}&=&\frac{\rho}{3}\sum_sx_s\left[
2\sigma_s+L_s+Q_s\left(L_s-\sigma_s\right)
(3\delta_{\beta z}-1)\right],\nonumber \\
\xi_{2\beta}&=&\frac{\rho}{3}\sum_s x_sv_s\left[
\frac{2}{\sigma_s}+\frac{1}{L_s}+Q_s\left(
\frac{1}{L_s}-\frac{1}{\sigma_s}\right)(3\delta_{\beta z}-1)\right],
\nonumber\\
\boldsymbol{\zeta}_2&=&
\left(\xi_{2y}\xi_{2z},\xi_{2z}\xi_{2x},\xi_{2x}\xi_{2y}\right),
\hspace*{3.3cm} (A.5) \nonumber
\end{eqnarray}
with $\beta=x,y,z$. 

The Fourier transform $\hat{C}_{s\mu,t\nu}({\bf q})$ of the direct 
correlation function  
has the same explicit expression as Eq. (\ref{lac}), except that 
the convolutions between different weights $\omega^{(\alpha)}_{s\mu}$ 
are to be substituted by the products of their corresponding Fourier transforms 
$\hat{\omega}^{(\alpha)}_{s\mu}$. These are 
\begin{eqnarray}
&&\hat{\omega}_{s\mu}^{(0)}({\bf q})=\prod_{\nu=1}^3\left[\cos\left(q_{\nu}
\sigma_{\mu\nu}^{(s)}/2\right)\right],\nonumber\\
&&\hat{\omega}_{s\mu}^{(3)}({\bf q})
=\prod_{\nu=1}^3 \left[
2\sin\left(q_{\nu}\sigma_{\mu\nu}^{(s)}/2\right)/q_{\nu}\right]\nonumber\\&&
\hat{\omega}_{s\mu}^{(1\alpha)}({\bf q})=\frac{2}{q_{\alpha}}
\tan\left(q_{\alpha}\sigma_{\mu\alpha}^{(s)}/2\right)
\hat{\omega}^{(0)}_{s\mu}({\bf q}),\nonumber\\&&
\hat{\omega}_{s\mu}^{(2\alpha)}({\bf q})=\frac{q_{\alpha}}{2}
\cot\left(q_{\alpha}\sigma_{\mu\alpha}^{(s)}/2\right)
\hat{\omega}^{(3)}_{s\mu}({\bf q}),
\setcounter{equation}{6}
\end{eqnarray}
where ${\bf q}=(q_1,q_2,q_3)$ is the weight vector, and 
$\sigma_{\mu\nu}^{(s)}=\sigma_s+\left(L_s-\sigma_s\right)\delta_{\mu\nu}$.


\begin{references}

\bibitem{Koda} T. Koda and H. Kimura, J. Phys. Soc. Jpn. {\bf 63}, 984 (1994).
\bibitem{vanRoij} R. van Roij and B. Mulder, 
Phys. Rev. E {\bf 54}, 6430 (1996)
\bibitem{Cinacchi} 
G. Cinacchi, E. Velasco and L. Mederos, J. Phys.: Condens. Matter {\bf 16}, 
S2003 (2004).
\bibitem{Cinacchi1}
G. Cinacchi, L. Mederos and E. Velasco, J. Chem. Phys. {\bf 121}, 3854 (2004).
\bibitem{Stroobants} A. Stroobants, Phys. Rev. Lett. {\bf 69}, 2388 (1992).
\bibitem{Bohle} A. M. Bohle, R. Holyst and T. Vilgis, Phys. Rev. Lett.
{\bf 76}, 1396 (1996).
\bibitem{Bates} M. A. Bates and D. Frenkel, J. Chem. Phys. {\bf 109}, 6193 (1998).
\bibitem{Parsons} J. D. Parsons, Phys. Rev. A {\bf 19}, 1225 (1979).
\bibitem{Lee} S.-D. Lee, J. Chem. Phys. {\bf 87}, 4972 (1987);
{\it ibid}. {\bf 89}, 7036 (1988).
\bibitem{Yuri} J. A. Cuesta, and Y. Mart\'{\i}nez-Rat\'on, 
Phys. Rev. Lett. {\bf 78}, 3691 (1997); J. Chem. Phys. {\bf 107}, 6379 (1997).
\bibitem{Yuri1} Y. Mart\'{\i}nez-Rat\'on, Phys. Rev. E {\bf 69}, 
061712 (2004).
\bibitem{Harrowell} A. Casey and P. Harrowell, J. Chem. Phys. {\bf 103}, 6143 (1995).
\bibitem{Somoza} A. M. Somoza and Tarazona, Phys. Rev. Lett. {\bf 61}, 2566
(1988); Phys. Rev. A {\bf 41}, 965 (1990).
\bibitem{depletion} 
Some predictions by Dogic et al. (Ref. 15) based on Onsager theory 
indicate that, for parallel hard rods, as their aspect ratio 
is increased beyond $\kappa_2\simeq 10$, the smectic phase should
stabilise at the expense of the nematic phase. This effect is related to 
a more favourable packing of the particles and therefore has an 
entropic origin (depletion interaction).
\bibitem{Dogic} 
Z. Dogic, D. Frenkel and S. Fraden, Phys. Rev. E {\bf 62}, 3925 (2000).
\end{references}
\end{document}